\documentstyle[preprint,tighten,aps]{revtex}
\begin{document}           %
\draft
\preprint{\vbox{\noindent
           \null\hfill CTP\#2360\\
          \null\hfill INFNCA-TH-94-24}}
\title{The Relevant Scale Parameter\\
           in the High Temperature Phase of QCD\\
      }

\author{Suzhou Huang$^{(1,2)}$\cite{email}
        and Marcello Lissia$^{(1,3)}$\cite{email} }
\address{
$^{(1)}$Center for Theoretical Physics, Laboratory for Nuclear Science
and Department of Physics, \\
Massachusetts Institute of Technology, Cambridge, Massachusetts 02139\\
$^{(2)}$Department of Physics, FM-15, University of Washington,
Seattle, Washington 98195~\cite{present}\\
$^{(3)}$Istituto Nazionale di Fisica Nucleare,
via Negri 18, I-09127 Cagliari, Italy~\cite{present}\\
and Dipartimento di Fisica dell'Universit\`a di Cagliari, I-09124
Cagliari, Italy
         }
\date{October 1994}
\maketitle                 
\begin{abstract}
We introduce the running coupling constant of QCD in the high
temperature phase, $\tilde{g}^2(T)$, through a renormalization
scheme where the dimensional reduction is optimal at the one-loop
level. We then calculate the relevant scale parameter, $\Lambda_T$,
which characterizes the running of $\tilde{g}^2(T)$ with $T$,
using the background field method in the static sector.
It is found that $\Lambda_T/\Lambda_{\overline{\text{MS}}}
=e^{(\gamma_E+1/22)}/(4\pi)\approx 0.148$. We further verify
that the coupling $\tilde{g}^2(T)$ is also optimal for lattice
perturbative calculations. Our result naturally explains why the
high temperature limit of QCD sets in at temperatures as low as
a few times the critical temperature. In addition, our $\Lambda_T$
agrees remarkably well with the scale parameter determined from
the lattice measurement of the spatial string tension of the SU(2)
gauge theory at high $T$.
\end{abstract}
\narrowtext
\section{Introduction}
\label{intro}
  At high temperatures QCD is expected to undergo a partial
dimensional reduction~\cite{dr-early,Landsman}, namely static
correlations at distances larger than the thermal wavelength
($1/T$) can be reproduced by a three dimensional Lagrangian,
where only the static modes of the original theory are present.
This reduced Lagrangian can be computed perturbatively up to
a specific order in the QCD running coupling constant. In fact,
non-perturbative infrared phenomena (e.g. thermo-mass generation)
prevent the complete reduction, i.e. reduction to all orders in the
QCD running coupling, from taking place~\cite{Landsman}. Consequently,
observables can be reproduced only up to corrections of a specific
order, before non-perturbative physics begins to dominate.

  Even though a complete dimensional reduction is not possible,
the partial dimensional reduction of QCD still provides a simplified
physical picture. However, phenomenological applications of this picture
depends crucially on how high is the temperature above which this
picture begins to take place. Since we expect corrections to vanish
with some power of the QCD coupling, and since at zero temperature
the asymptotic freedom starts dominating QCD physics at typical scales
of about 10 to 20 times $\Lambda_{\overline{\text{MS}}}$, one might
anticipate that the reduced theory should become effective only for
$T\gg\Lambda_{\overline{\text{MS}}}$.

  Contrary to this expectation, there are strong
evidences~\cite{reisz,bali,bs-lattice} that the dimensional reduction
picture is already valid at temperatures as low as two or three
times the critical temperature $T_c$ (the deconfining transition in
the pure Yang-Mills case or the chiral restoration in full QCD).
Since $T_c$ is numerically not very different from
$\Lambda_{\overline{\text{MS}}}$, and considering that the QCD
coupling constant only runs logarithmically, it is very surprising to
find that the high-$T$ regime of QCD starts at temperatures this low.

  This apparent puzzle can be solved with the observation that the
effective scale parameter for the reduced theory, $\Lambda_T$, is
renormalized after integrating out the non-static modes and becomes
drastically smaller than $\Lambda_{\overline{\text{MS}}}$. In fact,
the definition of a scale parameter that characterizes the approach
to the dimensional reduction regime implies the definition of a
suitable coupling constant, $\tilde{g}^2(T)$, that yields a sensible
perturbative expansion at high temperature, i.e. an expansion whose
coefficients contain minimal contribution from non-static modes.

  In this paper, we use the background field approach to define
and compute the relevant coupling constant, and hence the scale
parameter $\Lambda_T$. More specifically, we calculate the one-loop
effective action for the background field in the static sector, and
define the renormalization scheme by requiring that dimensional
reduction be optimal for this gauge-invariant quantity. Furthermore,
we verify by an explicit computation that this same renormalization
scheme is also optimal for lattice perturbative calculations at
high $T$, and therefore it provides a natural scale also for lattice
simulations.

  In section~\ref{optdr} we introduce the renormalization scheme that
defines the scale parameter $\Lambda_T$ within the background field
approach. In section~\ref{calc}, we apply this definition and calculate
$\Lambda_T / \Lambda_{\overline{\text{MS}}}$. First we perform the
calculation for the SU($N$) gauge theory in the continuum using
dimensional regularization; the effect of light quarks is also
considered. Then we repeat the calculation for the pure SU($N$) gauge
theory on the lattice in the Wilson formulation. In section~\ref{exp},
we compare our result with numerical determinations of $\Lambda_T$
from a lattice measurement of the spatial string tension at high $T$.
Section~\ref{conclu} is reserved for the conclusions. Several
technical points pertinent to the lattice perturbative calculation
at high $T$ are discussed in the Appendix.

\section{Dimensional Reduction and Optimal Renormalization Scheme}
\label{optdr}
  The standard SU($N$) Yang-Mills gauge theory reduces at the tree
level to the three dimensional Yang-Mills theory with adjoint Higgs
($\phi^a\equiv Q_0^a$)
\begin{equation}
{\cal L}_{\text{RD}}=-{1\over 4}F^a_{ij} F^a_{ij}
-{1\over 2}(D_i\phi)^a(D_i\phi)^a\, ,
\label{ym_3d}
\end{equation}
where $F_{ij}^a=\partial_i Q_j^a-\partial_j Q_i^a-g_3 f^{abc}
Q_i^b Q_j^c$ and $(D_i\phi)^a=\partial_i\phi^a-g_3 f^{abc}
Q_i^b\phi^c$. The coupling $g_3$ is related to the four dimensional
coupling through $g_3^2=g^2 T$. Since ${\cal L}_{\text{RD}}$ is a
super-renormalizable theory in three dimensions and there is no
other dimensionful scale around, all the dynamical scales must be
set by the coupling constant $g_3^2=g^2T$.

  Of course, once loop corrections are included the reduced theory
in Eq.~(\ref{ym_3d}) would acquire new vertices and the coupling
constant $g_3^2$ would depend on the original coupling $g^2$ in a
more complicated way. For example, $g_3^2$ would receive corrections,
such as $g^4T$ and so on. However, due to the asymptotic freedom of QCD
($g^2\sim 1/\ln T$) we still expect that dynamical scales
are set by $g_3^2\approx g^2T$, provided the scale parameter
is chosen in a suitable way.

  Therefore, we believe that the concept of dimensional reduction
involves two equally important aspects.
On one hand, there is the possibility of a simplified description
by using a theory ${\cal L}_{\text{RD}}$ with
less degrees of freedom in lower dimensions. On the other hand,
the evolution of the parameters of ${\cal L}_{\text{RD}}$ as a
function of temperature should be dictated by the original theory.
The main concern of our present work is to determine this evolution,
which in turn determines the temperature dependence of the relevant
physical observables.

\subsection{Background Field Method in the Static Sector}
  It is well known that the effective action calculated using
the background field method~\cite{abbott} is gauge invariant for the
background gauge field at $T=0$. This gauge invariance guarantees that
the coupling constant renormalization is related to the wavefunction
renormalization of the background field through $Z_g=Z_A^{-1/2}$.
Hence, the calculation of the quadratic part of the effective
action, i.e. the two-point function for the background field, is
sufficient to renormalize the coupling~\cite{abbott}.

  Moreover, to the leading order, there is no magnetic mass generation
at finite $T$. Therefore, the one-loop effective action for the magnetic
sector is invariant under time-independent gauge transformations also
at finite $T$, insuring that the relation $Z_g=Z_A^{-1/2}$ still holds
for the static background field
\begin{equation}
A^a_0(\tau,\bbox{x})=0,\,\,\,\,\, A^a_i(\tau,\bbox{x})
=A^a_i(\bbox{x})\, .
\label{bgf}
\end{equation}
The same conclusion can also be reached more formally by applying,
for instance, the methods of Ref.~\cite{abbott} to the background
field of Eq.~(\ref{bgf}). The residual gauge invariance in the static
sector implies that, in order to compute the coupling constant
renormalization at finite $T$, we only need to compute the two-point
function of the background field $A_i^a$ in the static sector.

We can still use the zero temperature Feynman rules, as given for
instance by Abbott~\cite{abbott}. The only difference in the
calculation is that time-components of all momenta become discrete
Matsubara frequencies $(2\pi nT)$, and the corresponding integrals
become discrete sums.

\subsection{Subtraction Scale}
As exhaustively discussed by Landsman~\cite{Landsman}, the
decoupling of the non-zero modes at high-$T$ is maximal only in
some specific renormalization schemes, such as the BPHZ scheme.
In the background field method we only need to fix one renormalization
condition: we demand that the two-point function for the background
field in the low external momentum (relative to $T$) limit coincides
with the contribution solely from zero modes.

Landsman~\cite{Landsman} uses a finite temperature renormalization
group approach, since he discusses thermal reduction in a more general
context where several couplings are present. Thanks also to the
background field approach, we deal with a simpler situation where
only one coupling needs renormalization.

Therefore, we can directly implement the renormalization condition by
using the freedom in the choice of the subtraction scale, $\mu$,
which becomes a function of $T$. Intuitively, we expect $\mu$ to be
of order of $T$. The purpose of our paper is to find out what is
the proportionality constant.

Then the reduced theory, Eq.~(\ref{ym_3d}), with the $T$-dependent
coupling $g_3^2=g^2(\mu(T))T$, reproduces the full two-point function
up to corrections of order of $\bbox{p}^2/T^2$ at the one-loop level.
Due to the gauge invariance, the two-point function for the static
background field $A^a_i$ must have the form
\begin{equation}
(\delta_{ij}\bbox{p}^2-p_i p_j)\,\delta_{ab}\,
\Pi_M(\bbox{p}^2,T,\mu)\, .
\end{equation}
Specifically, we choose $\mu$ by requiring the following
renormalization condition for the non-static contribution to
$\Pi_M(\bbox{p}^2,T,\mu)$:
\begin{equation}
\label{normal}
\Pi^{NS}_M(\bbox{p}^2=0,T,\mu(T))= 0 \, .
\end{equation}
The procedure is best explained by directly going through the
calculation in the next section.
\section{Calculation of the Scale Parameter}
\label{calc}
\subsection{In the Continuum}
In the continuum calculation we use dimensional regularization
in the spatial dimensions, that is
\begin{equation}
\int{d^4 k\over(2\pi)^4}\rightarrow
T\sum_{n=-\infty}^\infty \, \mu^{2\epsilon}
\int{d^{3-2\epsilon}\bbox{k}\over(2\pi)^{3-2\epsilon}}\, ,
\label{dim_reg}
\end{equation}
and the $\overline{\text{MS}}$ subtraction scheme.

At the one-loop level there are four graphs that contribute to
$\Pi_M$ in the full theory: the bubble and tadpole graphs for both
the quantum gauge fields and the ghost fields. The resulting total
contribution is
\begin{eqnarray}
\Pi_M(\bbox{p}^2,T,\mu)&=&{1\over g^2(\mu)}
-\biggl[{21\over 64}+{3\over 32}\alpha
+{1\over 64}\alpha^2\biggr] {N T\over\sqrt{\bbox{p}^2}} \nonumber \\
&-&\beta_0\biggl[\ln(\mu^2/T^2)+2\gamma_E-2\ln 4\pi
+{1\over 11} \biggr]
+{\cal O}(\bbox{p}^2/T^2)\, ,
\label{pi_m}
\end{eqnarray}
where $g^2(\mu)$ is the running coupling defined in the
$\overline{\text{MS}}$ scheme, $\beta_0=11N/(48\pi^2)$, $\alpha$
is the gauge parameter, and $\gamma_E$ is the Euler constant.

The first term in Eq.~(\ref{pi_m}) is obviously the classical
contribution. The second term is the contribution of the static
modes, and one can easily check that it can be reproduced by the
reduced theory, Eq.~(\ref{ym_3d}), with coupling constant
$g_3^2 = g^2(\mu)T$ and the same gauge parameter. The third
term is the one that must be eliminated according to our
renormalization prescription, which accomplishes maximal
decoupling~\cite{Landsman}. We obtain this result by choosing
\begin{equation}
\mu(T)=4\pi T e^{-(\gamma_E+c)}\, ,
\label{mu_t}
\end{equation}
where $c=1/22$. It is very reassuring to find that this optimal
choice of the subtraction scale $\mu$ is independent of the gauge
parameter $\alpha$. The remaining contributions in Eq.~(\ref{pi_m})
are suppressed by powers of $1/T^2$.

In summary, to achieve maximal decoupling and hence the optimal
dimensional reduction, the effective coupling in the reduced
theory must be
\begin{equation}
\tilde{g}^2(T)\equiv {1\over\beta_0\ln(T^2/\Lambda_T^2)}=
g^2(\mu)\bigg|_{\mu=4\pi T e^{-(\gamma_E+c)}}\, ,
\label{g_tilde}
\end{equation}
which defines the scale parameter
\begin{equation}
\Lambda_T={e^{(\gamma_E+c)}\over 4\pi}\Lambda_{\overline{\text{MS}}}\, .
\label{lambda_t}
\end{equation}

  This results has a clear physical interpretation. The non-static
modes decouple in the high-$T$ limit, but their presence is
nevertheless revealed by the appearance of the new scale $\Lambda_T$
in the reduced theory (without any reference to the original theory,
the only scale would be $T$). While this new scale is obviously related
to the scale $\Lambda_{\overline{\text{MS}}}$ that governs the full
theory at zero temperature, the two scales do not coincide. Only with
the coupling constant of Eq.~(\ref{g_tilde}), whose temperature
evolution is set by the scale $\Lambda_T$ in Eq.~(\ref{lambda_t}),
the reduced theory is capable of reproducing the full four-dimensional
one-loop corrections up to terms suppressed by $1/T^2$. Incidentally,
it is interesting to notice that, if one intuitively identifies
$2\pi T$ (rather than $T$) as the relevant frequency unit, one gets an
answer numerically close to the right one in Eq.~(\ref{mu_t}).

  At this point, we must point out that the optimal perturbative
dimensional reduction criterion alone does not uniquely determine
the scale $\Lambda_T$. In general, $\Lambda_T$ also depends on the
specific Green's functions for which we demand optimal reduction.
The use of a different process, represented by a different set of
Feynman graphs, yields a different $c$ in Eq.~(\ref{lambda_t}).
However, we believe that typically $|c|\lesssim 1$, and a different
choice should not modify the scale ratio in Eq.~(\ref{lambda_t})
in an essential way. For example, Landsman~\cite{Landsman} calculated
the temperature dependent coupling renormalization factor $Z_g$ by
imposing maximal dimensional reduction on the two- and three-point
functions in the conventional effective action (where the
relationship $Z_g=Z_A^{-1/2}$ no longer holds). He did not
express his result explicitly in terms of the scale ratio. But if
we do it, we find that his result is quite close to ours, i.e.
Eq.~(\ref{lambda_t}) with $c=0$.

  Another example that clearly shows the necessity of using an
optimal dimensional reduction scheme for defining the relevant scale
at high-$T$ can be found in the Gross-Neveu model~\cite{drl}. In that
model a similar strategy makes the sub-leading correction to the
screening mass of order of $\tilde{g}^6(T)$, rather than
$\tilde{g}^4(T)$, demonstrating that $\tilde{g}^2(T)$ is a sensible
expansion parameter.

  Of course, the optimal dimensional reduction criterion is not the
only way to define a temperature dependent coupling constant. For
example, the quark-antiquark potential at a distance of order of
$1/T$ is used to define $g^2(T)$ in Ref.~\cite{reisz}. While it is
certainly legitimate to make such a choice, it is also true that,
because the reduced theory is meant to reproduce the full theory
only at distances much larger than $1/T$ (spatial momenta small
compared to $T$), definitions of the couplings made by
matching short distance properties of the full
and reduced theories do not necessarily define a scale that correctly
characterizes the approach to the asymptotic high-$T$ regime.

  At last, let us consider the effect of quarks on our result.
If $N_f$ light quarks are present in the theory, results of
Eqs.~(\ref{mu_t}), (\ref{g_tilde}) and~(\ref{lambda_t}) still apply,
but with  $\beta_0=(11N-2N_f)/(48\pi^2)$ and
$c=(N/2-2N_f\ln4)/(11N-2N_f)$, where we have adopted the
convention for the trace of the Dirac-matrix:
$\text{Tr}\gamma_\mu\gamma_\nu=-4\delta_{\mu\nu}$.

  For the phenomenological relevant case of $N=N_f=3$, we get the
value $c=-0.2525$, which corresponds to
$\Lambda_T/\Lambda_{\overline{\text{MS}}}\approx 0.110$. Therefore,
additional flavors further decrease the scale ratio until
$N_f>16$ (for $N=3$), where asymptotic freedom is lost.

\subsection{On the Lattice}
\label{latt}
  The determination of the scale parameter $\Lambda_{T}$ that
governs the temperature dependence of the coupling in the reduced
theory only involves an one-loop calculation. Nevertheless, the
reduced theory is in general still non-perturbative, and
non-perturbative methods are necessary to extract information from
it. The standard approach is of course the lattice formulation.

  It should be clear from its definition (see section~\ref{optdr}
and Ref.~\cite{Landsman}) that the concept of maximal decoupling
scheme, along with the associated scale parameter, is independent of
how the theory is regularized, and we expect the same $\Lambda_{T}$
on the lattice, as long as we use the same renormalization condition,
Eq.~(\ref{normal}).

  On the other hand, since the lattice theory is usually defined in
terms of the bare lattice coupling, $g_0^2(a)$, without introducing
additional renormalization scale other than the lattice constant $a$,
it is quite interesting to see with an explicit calculation how this
scale emerges on the lattice at high $T$. In this respect, there
are close analogies between our choice of $\tilde{g}^2(T)$ as a
suitable expansion parameter for lattice perturbative calculation
at high temperature, and the necessity of using expansion parameters
different from the bare lattice coupling $g_0^2(a)$ for perturbative
calculations at zero temperature~\cite{lepage}.

  In the following we verify that optimal dimensional reduction for
the lattice effective action computed in the background field method
defines indeed the same scale parameter we have found in the continuum
calculation. For the sake of concreteness, we perform the calculation
for the pure SU($N$) Wilson action, but the same result is expected
to hold for other actions as well.

  In general, the coupling defined in the lattice background field
method should have the following dependence on the bare lattice
coupling up to one-loop
\begin{equation}
g_L^2(T)\equiv g_0^2(a)+g_0^4(a)\beta_0
\biggl[-\ln(a^2T^2)+c_L^T\biggr]\, .
\label{g_lattice}
\end{equation}
We want to show that $c_L^T$ is such that $g_L^2(T)=\tilde{g}^2(T)$.
Since we have expressed $\tilde{g}^2(T)$ in terms of $g^2(\mu)$ in
the $\overline{\text{MS}}$ scheme, see Eq.~(\ref{g_tilde}),
we use the known relation between $g_0^2(a)$ and $g^2(\mu)$ in
the $\overline{\text{MS}}$ scheme~\cite{lambda-lat,gonzalez}
\begin{equation}
g_0^2(a)=g^2(\mu)-
g^4(\mu)\beta_0\biggl[-\ln(a^2\mu^2)+c_L^0\biggr],
\end{equation}
and express also $g_L^2(T)$ in Eq.~(\ref{g_lattice}) in terms of
$g^2(\mu)$
\begin{equation}
g_L^2(T)=g^2(\mu)-g^4(\mu)\beta_0
\biggl[-\ln(\mu^2/T^2)-c_L^T+c_L^0\biggr]+{\cal O}(g^6(\mu))\, .
\label{g_lat2}
\end{equation}
By comparing Eq.~(\ref{g_tilde}) and Eq.~(\ref{g_lat2}) we see that
to show $g_L^2(T)=\tilde{g}^2(T)$ is equivalent to show that
\begin{equation}
c_L^T=c_L^0+2\gamma_E-2\ln(4\pi)+{1\over 11}\, ,
\label{c_lt}
\end{equation}
where the explicit expression of $c_L^0$, which has been calculated
by several authors~\cite{lambda-lat,gonzalez}, is
\begin{equation}
c_L^0 ={1\over 11}\biggl(
-11\gamma_E+2f_{11}+3f_{00}+6f_{10}
-1+24\pi^2 z_{10}+6\pi^2-6\pi^2/N^2\biggr)\, .
\label{delta}
\end{equation}
The constants $f_{ij}$ and $z_{ij}$ are defined as
\begin{equation}
f_{ij}\equiv (4\pi)^2 \int_0^\infty dx\, x\biggl[
e^{-8x}I_0^2(2x)I_i(2x)I_j(2x)-{\theta(x-1)\over(4\pi x)^2}\biggr]
\label{f_ij}
\end{equation}
and
\begin{equation}
z_{ij}\equiv \int_0^\infty dx\,
e^{-8x}I_0^2(2x)I_i(2x)I_j(2x)\, .
\label{z_ij}
\end{equation}
Here and in the following we have closely followed the notation
of Ref.~\cite{gonzalez}.

Since most of the calculation of the finite constant
$c_L^T$ is closely parallel to the calculation of $c_L^0$, we
only report the final result. In the Appendix, however, we
illustrate the only new ingredient that is not a trivial extension
of the calculation of $c_L^0$: the high temperature expansion on
the lattice. The lattice correspondent of the continuum result of
Eq.~(\ref{pi_m}) in the Feynman gauge ($\alpha=1$) is
\begin{equation}
\Pi_M^L(\bbox{p}^2,T,a)={1\over g_0^2(a)}-
{7\over 16} {NT\over\sqrt{\bbox{p}^2}}
-\beta_0 \biggl[-\ln(a^2 T^2)+c_L^T \biggr]
+{\cal O}\Bigl(\bbox{p}^2/T^2,a|\bbox{p}|,aT \Bigr)\, ,
\label{pil_m}
\end{equation}
with $c_L^T$ given by
\begin{equation}
c_L^T ={1\over 11}\biggl(
22\gamma_E+11\ln(4/\pi^2)+2f'_{11}+3f'_{00}+6f'_{10}
+24\pi^2 z_{10}+6\pi^2-6\pi^2/N^2\biggr)\, ,
\label{c_lt_cal}
\end{equation}
and
\begin{equation}
f'_{ij}\equiv (4\pi)^2 \int_0^\infty dx\,x\,e^{-2x}I_0(2x)
\biggl[e^{-6x}I_0(2x)I_i(2x)I_j(2x)-{1\over(4\pi x)^{3/2}}\biggr]\, .
\label{fp_ij}
\end{equation}
Since, as shown in the Appendix, $f'_{ij}=f_{ij}-\gamma_E-3\ln 4$,
this complete the proof of Eq.~(\ref{c_lt}) and, therefore, of the
fact that $g^2_L(T)=\tilde{g}^2(T)$.

  In other words, if we use $\tilde{g}^2(T)$ as the expansion parameter,
the lattice effective action in the high-$T$ limit takes the following
form
\begin{equation}
\Pi_M^L(\bbox{p}^2,T,a)={1\over \tilde{g}^2(T)}
-{7\over 16}{NT\over\sqrt{\bbox{p}^2}}
+{\cal O}\Bigl(\bbox{p}^2/T^2,a|\bbox{p}|,aT\Bigr)\, ,
\end{equation}
which is the same as its continuum counterpart, if we use
the same coupling $\tilde{g}^2(T)$ (see Eq.~(\ref{pi_m}) with
$\alpha=1$ and $\mu$ given by Eq.~(\ref{mu_t}) ).
In both cases we have been able to absorb in the coupling constant
all leading local corrections due to non-static modes, while the
non-local ones are reproduced by the reduced theory.

\section{Comparison to Lattice Result}
\label{exp}
  In the preceding section, we have demonstrated that $\Lambda_T$
is the relevant scale parameter in the high-$T$ limit. Our argument
is yet only perturbative in nature. However, as we emphasized earlier,
the determination of the scale parameter is largely one-loop effect.
Now let us compare our result with the scale parameter determined
from a non-perturbative method: lattice measurement of the spatial
string tension at high $T$. The primary reason for choosing the
spatial string tension~\cite{bali} rather than the heavy quark
potential at distances of order of $1/T$~\cite{reisz} is that the
concept of dimensional reduction only makes sense for large distance
(low momentum) quantities.

Bali et al.~\cite{bali} measured the spatial string tension
in SU(2) gauge theory as a function of temperature $\sigma_s(T)$.
Then they fitted their result to the expected form of the
string tension in the three-dimensional SU(2) Yang-Mills theory
\begin{equation}
\label{string}
 \sqrt{\sigma_s(T)} \propto g^2(T) T \, ,
\end{equation}
where the running of $g^2(T)$ with temperature is determined by
the SU(2) $\beta$-function
\begin{equation}
g^{-2}(T) = \frac{11}{12\pi^2}\ln(T/\Lambda_T)
+\frac{17}{44\pi^2}\ln\Bigl[2\ln(T/\Lambda_T)\Bigr].
\end{equation}
Even though the simulation process knows nothing about the dimensional
reduction, the fitting formula Eq.~(\ref{string}) in fact defines
the optimal three-dimensional coupling $g_3^2=g^2(T) T$ through the
string tension, similar in spirit to what we have done for the
background field effective action. As a result, their fitted value of
$\Lambda_T^{\sigma} = (0.076\pm 0.013)T_c$ is the first, to our
knowledge, non-perturbative determination of the scale that
characterizes the high-$T$ regime for the SU(2) gauge theory.

  In the scaling regime we expect that the critical temperature
behaves like
\begin{equation}
T_c={\Lambda_L\over N_\tau}
\biggl({11N^2\over 24\pi^2\beta_c}\biggr)^{51\over 121}
\, \exp\biggl({12\pi^2\over 11N^2}\beta_c\biggr)\, ,
\end{equation}
where $\beta_c=2N/g_0^2(a)$. From their critical coupling
$\beta_c=2.74$ at $N_\tau=16$, and the known ratio
$\Lambda_{\overline{\text{MS}}}/\Lambda_L=38.85\exp[-3\pi^2/(11N^2)]$,
it is straightforward to express $T_c$ in terms of
$\Lambda_{\overline{\text{MS}}}$:
$T_c=1.62\Lambda_{\overline{\text{MS}}}$.
Then their numerical measurement yields
\begin{equation}
\Lambda_T^{\sigma}=(0.123\pm 0.021)\Lambda_{\overline{\text{MS}}} \, ,
\end{equation}
which is remarkably close to our result
\begin{equation}
 \Lambda_T = \frac{e^{\gamma_E+1/22}}{4\pi}
             \Lambda_{\overline{\text{MS}}}
        \approx 0.148 \Lambda_{\overline{\text{MS}}} \, ,
\end{equation}
in spite of the different renormalization conditions.

  The smallness of
$\Lambda_T/T_c$
gives a natural explanation of why their spatial string tension already
at temperatures around $2T_c$ is numerically so close to the
string tension of the three dimensional SU(2) Yang-Mills gauge theory.

The result of Ref.~\cite{bali} implies a minor role for the Higgs
sector in the reduced theory, whereas the result of Ref.~\cite{reisz}
seemingly implies the contrary. It would be interesting to see whether
the use of an optimal coupling in calculations such as those in
Ref.~\cite{reisz} and focusing only on long distance quantities
could resolve the disagreement. For example, the gluonic
Debye-screening mass, $\mu_D\sim g(T)T$, could be used to define
yet another optimal coupling. It is rather unfortunate that the
numerical results in Refs.~\cite{reisz,gao} are not accurate enough
to determine a meaningful $\Lambda_T$.

\section{Conclusions}
\label{conclu}
  We have defined a temperature dependent running coupling constant
$\tilde{g}^2(T)$ in the spirit of the maximal decoupling of non-static
modes of Landsman~\cite{Landsman} for SU($N$) gauge theories. More
specifically, this coupling is such that the static modes reproduce
the quadratic part of the one-loop effective action for the background
field in the low momentum limit. Furthermore, $\tilde{g}^2(T)$
provides a meaningful expansion parameter in the high-$T$ limit, and
its dependence on temperature defines a typical scale $\Lambda_T$ for
the high-$T$ regime.

  We have calculated this coupling constant, and the related scale
parameter, first in the continuum with dimensional regularization,
where we verified its independence from the gauge fixing parameter
$\alpha$, and then showed that the same coupling is also optimal
for the lattice perturbative calculation at high $T$. Our results
are $\Lambda_T=0.148\Lambda_{\overline{\text{MS}}}$ in the pure
Yang-Mills or quenched cases and
$\Lambda_T=0.110\Lambda_{\overline{\text{MS}}}$ for $N=N_f=3$.

  We have argued that this scale is typical in the high-$T$ regime,
even if its precise value depends on the specific definition. The
consequence of our result is that the high-$T$ regime of QCD, where
the dimensional reduction picture appears to take place, sets in
at temperatures as low as a few times of the critical temperature.

  Our calculation is in very good agreement with the non-perturbative
determination of the scale parameter in the lattice
simulations~\cite{bali} in the SU(2) Yang-Mills theory, therefore
reinforcing the advantage of the renormalization scheme based
on the optimal dimensional reduction criterion.

  It would be of great interest to have other lattice measurements
of the scale parameter using other observables, such as the ones
related to the gluonic Debye-screening mass and the deviations of
the mesonic and baryonic screening masses from their free values.

  This work was supported in part by funds provided by the U.S.
Department of Energy (DOE) under contract number DE-FG06-88ER40427
and cooperative agreement DE-FC02-94ER40818.

\appendix
\section*{}

In this appendix we discuss several points of the high temperature
expansion in perturbative lattice calculations. First we use the ghost
bubble-graph to illustrate the general method, then we prove that
$f'_{ij}-f_{ij}=-\gamma_E-3\ln 4$, and finally discuss the convergence
of the frequency sums to the corresponding zero temperature integrals.

{}From the lattice action, see for instance Ref.~\cite{gonzalez},
we derive the following expression for the ghost bubble-graph
\begin{equation}
B_{\mu\nu}(\bbox{p})\equiv {N\over 4a^2\Omega}
\sum_k {(e^{-ik_\mu a}-e^{i(k_\mu-p_\mu)a})
(e^{-i(k_\nu-p_\nu)a}-e^{ik_\nu a})
\over \sum_\lambda(1-\cos k_\lambda a)
\sum_\rho(1-\cos(k_\rho-p_\rho)a)}\, ,
\end{equation}
where $\Omega$ is the space-time volume and $p=(0,\bbox{p})$.
Exponentiating the denominator and converting the spatial momentum
sums into integrals (we work in the infinite spatial volume limit),
we obtain
\begin{eqnarray}
B_{\mu\nu}(\bbox{p})&=&{N\over 4a^2N_\tau}\sum_{n=0}^{N_\tau-1}
\int_{-\pi}^\pi{d^3\bbox{k}\over(2\pi)^3}\int_0^\infty d\alpha d\beta\,
e^{-(\alpha+\beta)(4-\cos{2\pi n\over N_\tau})}\prod_{\lambda=1}^3
e^{\sqrt{\alpha^2+\beta^2+2\alpha\beta\cos p_\lambda a}
\cos(k_\lambda-\phi_\lambda)} \nonumber \\
& &\times
\biggl(e^{-i(k_\mu+k_\nu-p_\nu a)}+e^{i(k_\mu+k_\nu-p_\mu a)}
-e^{i(k_\mu-k_\nu-p_\mu a+p_\nu a)}-e^{-i(k_\mu-k_\nu )}\biggr)\, ,
\end{eqnarray}
where $\phi_\lambda$ is implicitly defined by
$\tan\phi_\lambda=\beta\sin(p_\lambda a)
/[\alpha+\beta\cos(p_\lambda a)]$.
Now we perform the spatial momentum integrals, yielding the modified
Bessel functions. For the sake of concreteness, let us consider the
component $\mu=1$ and $\nu=2$
\begin{eqnarray}
\label{b12p}
& &B'_{12}(\bbox{p})={N\over 4a^2N_\tau}
\sum_{n=1}^{N_\tau-1} \int_0^\infty d\alpha d\beta\,
e^{-(\alpha+\beta)(4-\cos{2\pi n\over N_\tau})}
\biggl[\prod_{\lambda=1}^2
I_1(\sqrt{\alpha^2+\beta^2+2\alpha\beta\cos p_\lambda a})\biggr]\\
& &\times I_0(\sqrt{\alpha^2+\beta^2+2\alpha\beta\cos p_3 a})
\biggl(e^{-i(\phi_1+\phi_2-p_2a)}+e^{i(\phi_1+\phi_2-p_1a)}
-e^{i(\phi_1-\phi_2-p_1a+p_2a)}-e^{-i(\phi_1-\phi_2)}\biggr)\, .
\nonumber
\end{eqnarray}
In Eq.~(\ref{b12p}) $B'$ is just $B$ without the $n=0$ term in the
frequency sum. This static term is in fact the one that is directly
reproduced by the reduced theory, and should be excluded from
the contribution due to non-static modes.

In the limit of $|\bbox{p}|a\ll 1$ and $|\bbox{p}|\ll T$ (we are
interested in the small lattice spacing and high-$T$ limit),
Eq.~(\ref{b12p}) further simplifies
\begin{equation}
\label{b12p2}
B'_{12}(\bbox{p})=-N{p_1 p_2\over 12}
\int_0^\infty dx\, x \biggl[{1\over N_\tau}\sum_{n=1}^{N_\tau-1}
e^{-x(1-\cos{2\pi n\over N_\tau})}\biggr]
e^{-3x}I_1^2(x)I_0(x)
+{\cal O}\Bigl(\bbox{p}^2/T^2,a|\bbox{p}|\Bigr)\, .
\end{equation}
The expression in Eq.~(\ref{b12p2}) diverges in the limit
$N_\tau\rightarrow\infty$. We explicitly isolate its divergent part with
the following subtraction
\begin{eqnarray}
B'_{12}(\bbox{p})&=&-N{p_1 p_2 \over 12}
\int_0^\infty dx\, x \biggl[{1\over N_\tau}\sum_{n=1}^{N_\tau-1}
e^{-x(1-\cos{2\pi n\over N_\tau})}\biggr]
\bigg\{e^{-3x}I_1^2(x)I_0(x)-{1\over(2\pi x)^{3/2}}\biggr\}\nonumber\\
&-&N {p_1 p_2 \over 12}\int_0^\infty dx\, x
\biggl[{1\over N_\tau}\sum_{n=1}^{N_\tau-1}
e^{-x(1-\cos{2\pi n\over N_\tau})}\biggr]{1\over(2\pi x)^{3/2}}
+{\cal O}\Bigl(\bbox{p}^2/T^2,a|\bbox{p}|\Bigr)\, .
\end{eqnarray}
Now the first term is finite in the limit $N_\tau\rightarrow\infty$,
i.e. the limit  $a\rightarrow 0$ with $aN_\tau=1/T$ fixed, and it is
equal to $-N p_1 p_2 f'_{11}/(48\pi^2)$. We then use in the second
term the expansion
\begin{equation}
{\pi\over N_\tau}\sum_{n=1}^{N_\tau-1}{1\over\sin(\pi n/N_\tau)}
=2\gamma_E+\ln\biggl({4\over\pi^2 a^2T^2}\biggr)+{\cal O}(aT)\, ,
\end{equation}
and obtain
\begin{equation}
B'_{12}(\bbox{p}) = -N\frac{p_1 p_2 }{48\pi^2}
\Bigl[f'_{11}+2\gamma_E+\ln(4/\pi^2)-\ln(a^2 T^2)\Bigr]
+{\cal O}\Bigl(\bbox{p}^2/T^2,a|\bbox{p}|,aT \Bigr)\, .
\end{equation}

Next we want to relate the finite integrals that are found in the
lattice high-$T$ expansion $f'_{ij}$, defined in Eq.~(\ref{fp_ij}),
to the corresponding integrals that are found in
the zero temperature calculation $f_{ij}$, defined in Eq.~(\ref{f_ij}).
Directly from their definitions, we find
\begin{eqnarray}
f'_{ij}-f_{ij}
&=& \lim_{\epsilon\rightarrow 0}
\int_0^\infty dx\,e^{-\epsilon x}\biggl[{\theta(x-2)\over x}
-\sqrt{2\pi}{e^{-x}I_0(x)\over\sqrt{x}}\biggr] \\
&=& \lim_{\epsilon\rightarrow 0} \,
\Bigl[-Ei(-2\epsilon) - 2Q_{-1/2}(1+\epsilon)\Bigr]   \, .
\end{eqnarray}
The expression is finite, and we have introduced a convergence factor
$e^{-\epsilon x}$ in the integral only to be able to integrate
separately the two terms. At last we obtain the desired result
$f'_{ij}-f_{ij}=-\gamma_E-3\ln 4$
by using the small epsilon expansions of the exponential-integral
function $Ei(-2\epsilon)=\gamma_E+\ln(2\epsilon)+{\cal O}(\epsilon)$
and of the Legendre function of the second kind
$2Q_{-1/2}(1+\epsilon)=-\ln(2\epsilon)+3\ln 4+{\cal O}(\epsilon)$.

  The last issue we would like to address in this appendix is the
high-$T$ expansion of those terms independent of external momentum,
such as the tadpole graphs. Physically we expect that these terms
cannot contain $\ln(aT)$, since they do not contribute to the
renormalization, and in fact they should eventually cancel out due
to the gauge invariance or the lack of magnetic mass generation at
the one-loop level. Therefore we should be able to factorize any
power dependence on $T$ trivially, and take the continuum limit of
the frequency sums. Mathematically, this is guaranteed by the fact
that the convergence of the limit
\begin{equation}
\lim_{N_\tau\rightarrow \infty}
\frac{1}{N_\tau}\sum_{n=0}^{N_\tau-1}
f\left(\cos{\frac{2\pi n}{N_\tau}}\right)
= \int_{0}^{1} dx\, f\left(\cos{2\pi x}\right)
\end{equation}
is exponential, at least when $f(z)$ can be expanded as a power
series in $z$, which includes the cases we are concerned with.
Note that the terms with $n=0$ should be included in these
tadpole-like graphs, since they are not reproducible by the
reduced theory.

\end{document}